\documentclass[prb,twocolumn,amsmath,amssymb,nofootinbib,floatfix]{revtex4}
\usepackage{graphicx}
\usepackage{amsmath}
\usepackage{amssymb}
\usepackage{bm}
\usepackage{color}
\usepackage{hyperref}
\usepackage{amsthm}
\usepackage{booktabs}

\def\bea{\begin{eqnarray}}
\def\eea{\end{eqnarray}}
\def\<{\langle}
\def\>{\rangle}

\def\Tr{\mathrm{Tr}}
\def\be{\begin{equation}}
\def\ee{\end{equation}}

\def\ep{\varepsilon}
\def\avg#1{\overline{#1}}
\def\Id{\mathbf{1}}
\def\dd{\mathrm{d}}
\def\om{\omega}
\def\Del{\Delta}
\def\adag{a^\dagger}
\def\nbar{\bar n}
\def\Ntr{N_{\rm tr}}
\def\epc{\avg{\mathrm{ep}}_{\rm coh}}
\def\epb{\avg{\mathrm{ep}}}
\def\sx{\sigma_x}
\def\sy{\sigma_y}
\def\sz{\sigma_z}
\def\rvec{\mathbf r}
\begin{document}

\setcitestyle{square,numbers,comma}
\makeatletter
\def\@biblabel#1{[#1]}
\makeatother

\title{Entangling Power and  Symmetries in the Quantum Rabi Model}

\author{Ian Low$^{\,a,b}$, Jens Koch$\,^a$, Sahel Ashhab$^{\,c,d}$ }
\affiliation{\vspace{0.2cm}
\mbox{$^a$Department of Physics and Astronomy, Northwestern University, Evanston, IL 60208, USA}\\
\mbox{$^b$ High Energy Physics Division, Argonne National Laboratory, Lemont, IL 60439, USA}\\
\mbox{$^c$Advanced ICT Research Institute, National Institute of Information and Communications Technology,}\\
\mbox{4-2-1 Nukui-Kitamachi, Koganei, Tokyo 184-8795, Japan}\\
\mbox{$^d$Research Institute for Science and Technology, Tokyo University of Science,}\\
\mbox{1-3 Kagurazaka, Shinjuku-ku, Tokyo 162-8601, Japan}
}
\begin{abstract}
The quantum Rabi model is a standard effective Hamiltonian in studies of
light--matter interaction, capturing the simplest nontrivial setting in which a
qubit couples to a single harmonic oscillator.  Within the broader Rabi family, we focus on two special cases:
the Jaynes--Cummings (JC) model, which carries an explicit $U(1)$ symmetry, and
the asymmetric quantum Rabi model (AQRM), which possesses a parameter-dependent
``hidden'' symmetry that appears only at integer bias,
$\ep/\om\in\mathbb{Z}$, and is not manifest in the Hamiltonian.  We use the
time-averaged entangling power as an operator-level diagnostic of these
symmetry structures.  Since the oscillator Hilbert space is infinite-dimensional,
we compare two finite input ensembles: a Haar average after Fock-space
truncation and a coherent-state average at fixed mean occupation $\nbar$.
Both diagnostics show peaks at the integer-bias points of the AQRM, where the hidden
symmetry resides. In contrast, the manifest $U(1)$ symmetry at the JC point
instead gives a weak dip. Thus, the time-averaged entangling power responds to both
the hidden symmetry and the $U(1)$ symmetry in the Rabi family, with the sign
of the response indicating how the symmetry reorganizes the spectral expansion. These results demonstrate that the entangling power can serve as an operator diagnostic to reveal the presence and properties of hidden and manifest symmetries in light-matter systems.
\end{abstract}

\maketitle

\section{Introduction}
\label{sec:intro}

The quantum Rabi model (QRM) is the standard effective Hamiltonian used in a vast literature on light--matter interaction, describing a single qubit coupled to a single quantized harmonic mode,
\be
H=\om\left(\adag a+\tfrac12\right)+\frac{\Del}{2}\sz+\frac{\ep}{2}\sx+g\,\sx(a+\adag),
\label{eq:H_aqrm}
\ee
where $a$ and $\adag$ annihilate and create photons in the oscillator mode, respectively, with $\om$ being the oscillator frequency. In addition, $\sx$ and $\sz$ are qubit Pauli operators, $\Del$ is the qubit gap, $\ep$ is the qubit transverse bias, and $g$ is the qubit--oscillator coupling strength; we set  $\hbar=1$ throughout. Descending from Rabi's semiclassical analysis of a spin in an oscillating field~\cite{Rabi:1937}, the quantized model resisted closed-form solution for decades, until Braak established its analytic solvability from the discrete parity symmetry present at $\ep=0$~\cite{Braak:2011aa} (see Ref.~\cite{Xie:2017} for a review). Beyond its role as a solvable model, Eq.~(\ref{eq:H_aqrm}) underlies much of cavity and circuit quantum electrodynamics. Superconducting circuits now reach the ultrastrong- and deep-strong-coupling regimes, where $g$ rivals or exceeds $\om$, and the non-number-conserving part of the light--matter interaction can no longer be neglected~\cite{FornDiaz:2010bss,Niemczyk:2010,Yoshihara:2017,Yoshihara:2017pra,FornDiaz:2018,FriskKockum:2019,Blais:2021}. Trapped ions can also be used to simulate the same Hamiltonian across coupling regimes~\cite{Pedernales:2015,Lv:2018}, and a suitable scaling limit realizes a single-qubit quantum phase transition observed experimentally~\cite{Bakemeier:2012qpt,Ashhab:2013qpt,Hwang:2015,Cai:2021}. As an effective Hamiltonian, Eq.~(\ref{eq:H_aqrm}) is especially useful because its parameters are set by the underlying device and, in several platforms, can be tuned independently~\cite{Song:2026}.

Two limits organize the broader Rabi family. Writing $\sigma_\pm=(\sx\pm i\sy)/2$, the coupling $g\,\sx(a+\adag)$ contains both the excitation-number-conserving terms $\sigma_+a+\sigma_-a^\dagger$ and the counter-rotating terms $\sigma_+a^\dagger+\sigma_-a$. The rotating-wave approximation (RWA) drops the latter, giving the Jaynes--Cummings (JC) model, in which the excitation number $\adag a+\sigma_+\sigma_-$ is conserved~\cite{Jaynes:1963zz}; this manifest $U(1)$ symmetry makes it block-solvable and underlies much of quantum optics~\cite{ShoreKnight:1993}. The anisotropic generalization assigns independent couplings to these two pieces, $g\,(\sigma_+a+\sigma_-a^\dagger)+g'\,(\sigma_+a^\dagger+\sigma_-a)$. It interpolates between the JC point at $g'=0$ and the standard QRM at $g'=g$, where the coupling returns to $g\,\sx(a+\adag)$ and, at zero bias, the only visible symmetry is the discrete parity. Along with the transverse bias, both couplings can be tuned in a superconducting device, spanning the JC and anti-JC regimes~\cite{Song:2026}. The family thus realizes a continuous, a discrete, and a hidden symmetry within one Hamiltonian.

The asymmetric quantum Rabi model (AQRM) is defined by Eq.~(\ref{eq:H_aqrm}) with nonzero bias $\ep$, the offset from the symmetry point; it breaks the parity of the symmetric model and is routinely tunable in flux-qubit circuits. The biased Hamiltonian shows no manifest symmetry, yet exact level crossings occur in the AQRM in the integer-bias sectors, $\ep/\om\in\mathbb{Z}$. These were first traced to the exceptional (``Juddian'') spectrum through the divisibility of constraint polynomials~\cite{LiBatchelor:2015,Wakayama:2017,Kimoto:2020}. One of us then gave numerical evidence that clarified why it is difficult to unveil the hidden symmetry: the partition of states into symmetry classes depends on the model parameters~\cite{Ashhab:2019xft}. Mangazeev, Batchelor, and Bazhanov then constructed, for each integer bias, an operator $J_k$ commuting with the Hamiltonian, establishing the symmetry at the operator level~\cite{Mangazeev:2020yjs}; further properties followed~\cite{ReyesBustos:2021}, while its microscopic origin remains under study~\cite{Yang:2025}. It is worth stressing a point that will be important below: at generic fixed $(g,\Del)$ the integer-bias condition does not by itself imply an eigenvalue degeneracy. Exact crossings occur only at specific values of $(g,\Del)$, so the hidden symmetry leaves no generic spectral-degeneracy signature in a bias scan.

The entangling power of a unitary operator measures the entanglement it generates from product inputs, averaged over those inputs~\cite{Zanardi:2000zz,Zanardi:2001qu}. It is a property of the operator rather than of any single state, and it requires no prior knowledge of the system's symmetries; applied to the time-evolution operator and averaged over time, it characterizes the typical entanglement production of a dynamics~\cite{Lu:2008zz,Pal:2018epsc,Pal:2024xlp}. A growing body of work in particle and nuclear physics has found that the entanglement generated in scattering is extremized at points of enhanced symmetry: the nucleon--nucleon $S$-matrix is minimally entangling at the Wigner $SU(4)$ symmetric point~\cite{Beane:2018oxh,Low:2021ufv,Low:2026evp}, with similar correlations in low-energy hadron dynamics~\cite{Liu:2022grf,Liu:2023bnr,Hu:2025lua}, in extensions of the Higgs sector~\cite{Carena:2023vjc,Carena:2025wyh,Li:2026chm}, and in the general $S$-matrix framework~\cite{Cervera-Lierta:2017tdt,McGinnis:2025brt,McGinnis:2025xgt}. These observations have motivated the broader possibility that symmetry may be tied to an entanglement-extremization principle, and that the extremization of quantum resources may help organize fundamental parameters, from flavor structure~\cite{Thaler:2024anb} to the Higgs mass~\cite{Liu:2025iwh}. A closely related line of work replaces entanglement by nonstabilizerness, or ``magic,'' and asks whether this quantum resource may also organize physical structure, including in the weak mixing angle~\cite{Liu:2025bgw,Cao:2026aye}, a fundamental constant of nature, and in other settings~\cite{Liu:2025qfl,Liu:2025frx,Li:2026cpc,Busoni:2025dns,Gargalionis:2025iqs}.

In quantum many-body physics, the interplay of entanglement and symmetry has long been studied through state-level diagnostics---ground-state entanglement entropy and its area laws~\cite{Calabrese:2004eu,Calabrese:2009qy,Vidal:2002rm,Latorre:2003kg,Amico:2007ag,Eisert:2008ur}, and symmetry-resolved entanglement~\cite{Goldstein:2017bua,Xavier:2018eua,Murciano:2020lqq,Ares:2022koq}---which characterize a particular state and presuppose knowledge of the conserved charges.  Studies of entanglement of particular states in the AQRM also exist~\cite{Shi:2021}.

The time-averaged entangling power is complementary: a property of the dynamics that needs no such input, it dips at points of enhanced symmetry and at selected integrable points in anisotropic Heisenberg spin chains~\cite{Low:2026ep}. These programs share a common setting: finite-dimensional Hilbert spaces---of discrete quantum numbers or of spins---and symmetries that manifest as degeneracies. The QRM is neither: it couples a qubit to an oscillator whose Hilbert space is infinite-dimensional, and its hidden symmetry leaves no generic degeneracy. Whether an operator-level entanglement diagnostic can detect such a symmetry is, to our knowledge, untested.  

In this work we use the time-averaged entangling power as an operator-level diagnostic of symmetry in the Rabi family.  For the oscillator, there is no normalizable Haar measure on the full Hilbert space.  One may Haar-average after projecting onto a finite Fock space, obtaining a finite-Haar diagnostic whose typical input occupation is set by the cutoff.  Alternatively, one may use coherent states at a chosen input scale $\nbar$; in the main text we use fixed-radius coherent sampling, and in Appendix~\ref{app:coherent} we compare it with Gaussian coherent sampling at the same $\nbar$.  The finite-Haar and coherent diagnostics both reveal peaks at the integer-bias points where the hidden symmetry resides.

The remainder of this paper is organized as follows. Section~\ref{sec:qrm_family} briefly reviews the quantum Rabi model family and the symmetry structures relevant for this work. Section~\ref{sec:ep_coherent} defines the coherent-state entangling power and contrasts it with finite-Haar diagnostics. Section~\ref{sec:aqrm_peaks} presents the integer-bias peaks and derives their locations. Section~\ref{sec:sign} explains why the integer feature is a peak rather than a dip. Section~\ref{sec:jc} gives related results for the JC model, and Sec.~\ref{sec:discussion} concludes. Technical details are collected in the appendices.

\section{The Quantum Rabi Model Family}
\label{sec:qrm_family}

We next briefly review the quantum Rabi model family, emphasizing the symmetry structures that will be probed by the entangling-power diagnostic. In the convention of Eq.~(\ref{eq:H_aqrm}), the symmetric QRM is obtained by setting $\ep=0$. It has a ${\mathbb Z}_2$ parity symmetry represented by the operator
\be
\Pi=\sz\,(-1)^{\adag a}.
\label{eq:qrm_parity}
\ee
Indeed, $(-1)^{\adag a}$ changes the sign of $a+\adag$, while conjugation by $\sz$ changes the sign of $\sx$. Their product therefore leaves the interaction Hamiltonian $g\,\sx(a+\adag)$ invariant, and one can easily verify that $[\Pi,H]=0$ when $\ep=0$. The transverse bias term $\ep\sx/2$, on the other hand, changes sign under $\Pi$ and breaks this easily identifiable ${\mathbb Z}_2$ symmetry. This is the first distinction that will matter below: the symmetric QRM has a manifest discrete symmetry, whereas the AQRM at nonzero bias does not.

It is useful to recall how the JC model sits inside the same family. The light--matter coupling Hamiltonian can be decomposed as
\be
g\,\sx(a+\adag)
=
g(\sigma_+a+\sigma_-\adag)
+
g(\sigma_+\adag+\sigma_-a).
\label{eq:rotating_counterrotating}
\ee
The first term conserves the excitation number $\hat N=\adag a+\sigma_+\sigma_-$, while the second term changes it by two units. When the qubit and oscillator frequencies are close, $|\Del-\om|\ll \Del+\om$, and the coupling is weak, $|g|\ll \Del+\om$, the counter-rotating term oscillates rapidly in the interaction picture and is dropped in the RWA. This gives the JC model, whose conserved $\hat N$ generates a manifest $U(1)$ symmetry and decomposes the Hilbert space into finite-dimensional excitation sectors~\cite{Jaynes:1963zz,ShoreKnight:1993}. In contrast, the full QRM retains both terms in Eq.~(\ref{eq:rotating_counterrotating}); excitation number is not conserved, and the only symmetry at $\ep=0$ is the parity described by Eq.~(\ref{eq:qrm_parity}).

The anisotropic Rabi model makes the interpolation between the QRM and JC model explicit by allowing the rotating and counter-rotating couplings to be different,
\begin{align}
H_{\rm aniso}
=&
\om\left(\adag a+\tfrac12\right)
+\frac{\Del}{2}\sz
+\frac{\ep}{2}\sx
\nonumber\\
&+g_r(\sigma_+a+\sigma_-\adag)
+g_{\rm cr}(\sigma_+\adag+\sigma_-a).
\label{eq:anisotropic_qrm}
\end{align}
The JC point is $g_{\rm cr}=0$, the anti-JC point is $g_r=0$, and the standard QRM corresponds to $g_r=g_{\rm cr}$. Together with the bias $\ep$, these tunable parameters give a single framework containing the JC, anti-JC, symmetric-QRM, and biased-AQRM regimes~\cite{Song:2026}.

The AQRM is obtained by turning on the bias $\ep$ in Eq.~(\ref{eq:H_aqrm}). Since the parity in Eq.~(\ref{eq:qrm_parity}) is broken, the Hamiltonian has no easily visible symmetry of the QRM type. Nevertheless, at the commensurate values
\[
\ep/\om=k,\qquad k\in{\mathbb Z},
\]
the model has a hidden symmetry: the spectrum contains exceptional crossings, and, more generally, the Hilbert space can be organized into symmetry sectors whose assignment depends on the Hamiltonian parameters~\cite{LiBatchelor:2015,Wakayama:2017,Kimoto:2020,Ashhab:2019xft}. It is this dependence on the exact system parameters that makes the symmetry hidden.

Let us spell out the operator construction, since it is the most concrete sense in which the symmetry exists. Mangazeev, Batchelor, and Bazhanov use the convention
\be
\tilde H=\adag a+\tilde g\,\sx(a+\adag)+\tilde\Delta\,\sz+\tilde\epsilon\,\sx,
\label{eq:H_mbb}
\ee
which is related to Eq.~(\ref{eq:H_aqrm}) by $\tilde H=H/\om-1/2$, $\tilde g=g/\om$, $\tilde\Delta=\Del/(2\om)$, and $\tilde\epsilon=\ep/(2\om)$. They search for a self-adjoint operator that, in the $\sz$ basis, has the block form
\be
J_k= {\cal P}
\begin{pmatrix}
A_k(\adag,a) & B_k(\adag,a)\\
C_k(\adag,a) & D_k(\adag,a)
\end{pmatrix},
\qquad
{\cal P}=(-1)^{\adag a}.
\label{eq:Jk_ansatz}
\ee
Here the displayed matrix acts on the qubit space, while each entry is an operator on the oscillator Hilbert space. The factor ${\cal P}$ is the oscillator parity and is understood to multiply each normally ordered oscillator operator from the left.
Substituting this ansatz into $[J_k,H]=0$ gives recurrence relations, or equivalently generating-function equations, for the coefficients. After removing the trivial freedom to add to $J_k$ polynomials   in $H$,\footnote{The condition $[J_k,H]=0$ determines $J_k$ only up to a polynomial in $H$, since any polynomial in $H$ commutes with $H$.} nontrivial solutions occur at $\tilde\epsilon=k/2$, namely at $\ep=k\om$ in our convention~\cite{Mangazeev:2020yjs}.

The first nontrivial case already illustrates why the symmetry is hidden. At $\ep=\om$, define $\gamma=g/\om$ and $\eta=\Del/(2g)$. Starting from Eq.~(\ref{eq:Jk_ansatz}) with entries linear in $a$ and $\adag$, the commutator condition $[J_1,H]=0$ fixes the coefficients up to an overall normalization and the freedom to add a polynomial in $H$~\cite{Mangazeev:2020yjs}. In the $\sz$ basis, one may choose
\be
J_1={\cal P}
\begin{pmatrix}
\adag-a+2\gamma+\eta & \adag+a\\
-(\adag+a) & -(\adag-a)-2\gamma+\eta
\end{pmatrix}.
\label{eq:J1_hidden}
\ee
One can directly verify that $[J_1,H]=0$ at $\ep=\om$. The entries depend explicitly on the Hamiltonian parameters, so the symmetry sectors are not fixed once and for all as in the parity-symmetric QRM.

This should be contrasted with the symmetries more familiar from the JC model or from ordinary global symmetries. In the JC limit the conserved charge $\hat N$ is simple, parameter independent, and visible from the Hamiltonian; once the RWA has been made, the same operator organizes the Hilbert space into fixed excitation sectors for all values of the remaining system parameters. The same is true, in spirit, for a usual global symmetry: the generator is specified independently of the Hamiltonian parameters, the representation content of the Hilbert space is fixed, and the Hamiltonian is constrained by requiring it to commute with that generator. The AQRM hidden symmetry works in the opposite direction. The Hamiltonian is given first, the special bias value is then selected, and only at that point does one construct a parameter-dependent operator $J_k$ that commutes with it. Moreover, $J_k$ is not an involutive parity operator with $J_k^2=1$, but an algebraic object whose square is a polynomial in the shifted Hamiltonian. Thus the ``symmetry sectors'' are not fixed charges known in advance; they depend on $(g,\Del)$ and on the integer $k$. This is why the symmetry is hidden in a stronger sense than merely being written in an inconvenient basis.

For higher integer bias, the same construction gives operators $J_k$ whose local matrix entries have increasing degree in $a$ and $\adag$. In this normalization, $J_k^2$ is a polynomial of degree $k$ in the shifted Hamiltonian. Further structural properties of this commutant were proven in Ref.~\cite{ReyesBustos:2021}, while its microscopic origin continues to be investigated~\cite{Yang:2025}.

A final comment is important for the interpretation of our results. In the JC model, symmetry is visible as a conserved excitation number and, after a finite Fock-space truncation, as the block and frequency-degeneracy structure responsible for the finite-Haar dip. In the AQRM, by contrast, the hidden-symmetry condition $\ep/\om\in{\mathbb Z}$ does not by itself imply that a generic bias scan at fixed $(g,\Del)$ passes through an exact level crossing. The Juddian crossings occur only on special parameter loci. Thus the question addressed in this work is not whether the entangling power rediscovers an obvious degeneracy, but whether an operator-level dynamical diagnostic can respond to a symmetry whose organizing operator is both nonmanifest and parameter dependent.

\section{Entangling Power and Coherent-State Sampling}
\label{sec:ep_coherent}

We now define the entanglement diagnostic used in the rest of this paper.  For a QRM system, it is useful to distinguish the entangling power from the entanglement of an eigenstate, a ground state, or a thermal state.  One first prepares an unentangled bipartite input, evolves it with a unitary operator $U$, and asks how entangled the two subsystems become.  The result is then averaged over an ensemble of product inputs.  In this sense the entangling power is a property of $U$, and for $U(t)=\exp(-iHt)$ it is a property of the dynamics generated by $H$.

Let us begin with the standard finite-dimensional definition of the entangling power.  For a bipartite unitary $U$ acting on ${\cal H}_A\otimes{\cal H}_B$, take a product input $|\psi,\phi\>\equiv|\psi\>_A\otimes|\phi\>_B$.  The output reduced density matrix of subsystem $A$ is
\be
\rho_A=\Tr_B\left(U|\psi,\phi\>\<\psi,\phi|U^\dagger\right).
\label{eq:rhoA_def}
\ee
We measure the entanglement of the pure output state by the linear entropy
\be
{\cal E}_A(U;\psi,\phi)\equiv 1-\Tr\rho_A^2 ,
\label{eq:linear_entropy}
\ee
which vanishes when the output is a product state and is maximal for a maximally entangled pure output state. The Zanardi--Zalka--Faoro entangling power is the Haar average of Eq.~(\ref{eq:linear_entropy}) over independent product inputs~\cite{Zanardi:2000zz,Zanardi:2001qu},
\be
\mathrm{ep}(U)=\frac{d_A}{d_A-1}\int\!\dd\mu_{\rm H}(\psi)\int\!\dd\mu_{\rm H}(\phi)\,
\left[1-\Tr \rho_A^2\right],
\label{eq:ep_haar}
\ee
where $d_A=\dim{\cal H}_A$.  The prefactor normalizes the linear entropy so that each output state contributes a number between 0 and 1: it vanishes when $\rho_A$ is pure and equals unity when $\rho_A=\Id_A/d_A$ is maximally mixed.  Thus $0\le \mathrm{ep}(U)\le 1$.  In the AQRM, we will take subsystem $A$ to be the qubit, so $d_A=2$ and the prefactor is simply $2$.

For a two-qubit system, the only unitary operators with vanishing entangling power are, up to single qubit unitaries, the Identity and the SWAP gates~\cite{Low:2021ufv}. It is interesting to mention in passing here that this distinction has a direct realization in low-energy neutron--proton scattering: the $S$ matrix is minimally entangling as the Identity when Wigner's $SU(4)$ spin-flavor symmetry is realized, and as the SWAP when the dynamics exhibits nonrelativistic conformal invariance~\cite{Low:2021ufv}.

The use of the linear entropy above allows one to compute the entangling power algebraically, without explicitly performing the Haar averages~\cite{Zanardi:2000zz,Zanardi:2001qu,Lu:2008zz}. Since $\Tr\rho_A^2$ contains two copies of the reduced density matrix, the Haar input average can be recast as traces of SWAP operators on a doubled Hilbert space.  This is the standard swap formalism behind the entangling power; we review the details in Appendix~\ref{app:swap_formalism}.

The entangling power of the time-evolution operator is a time-dependent quantity. It is then natural to define the time-averaged entangling power
\be
\epb\equiv
\lim_{T\to\infty}\frac{1}{T}\int_0^T\!\dd t\,
\mathrm{ep}\!\left(e^{-iHt}\right) \ ,
\label{eq:ep_haar_time}
\ee
which is independent of $t$ and is therefore a property of the Hamiltonian $H$.\footnote{In high-energy scattering, we compute the entangling power of the $S$-matrix, 
which maps asymptotic in-states to asymptotic out-states and no additional time average is involved.} It is useful to spell out what the time average does before specializing to any particular input ensemble.  
Consider the product state $|\psi,\phi\>$ in the finite-dimensional Haar case above.  Write $|\psi,\phi\>=\sum_m c_m |E_m\>$ in the energy basis of $H$.  Choose a product basis $\{|a,b\>\}$ and define the $d_A\times d_B$ coefficient matrix of the eigenstate $|E_m\>$ by
\be
\begin{split}
|E_m\>&=\sum_{a=1}^{d_A}\sum_{b=1}^{d_B}(C_m)_{ab}|a,b\>,\\
(C_m)_{ab}&=\<a,b|E_m\>.
\end{split}
\label{eq:Cm_def}
\ee
It is useful to define the resulting operator on ${\cal H}_A$ by
\be
M_{mn}\equiv \Tr_B |E_m\>\<E_n| .
\label{eq:Mmn_def}
\ee
In components, the partial trace contracts the subsystem-$B$ index,
\begin{align}
(M_{mn})_{aa'}
&=
\sum_{b=1}^{d_B}
(C_m)_{ab}(C_n)^*_{a'b}
\nonumber\\
&=
(C_m C_n^\dagger)_{aa'} .
\label{eq:Mmn_components}
\end{align}
Thus $M_{mn}=C_mC_n^\dagger$ is a $d_A\times d_A$ matrix acting on subsystem $A$, and the reduced density matrix at time $t$ is
\be
\rho_A(t)=
\sum_{m,n}c_m c_n^* e^{-i(E_m-E_n)t} M_{mn}.
\label{eq:rhoA_spectral}
\ee
The advantage of Eq.~(\ref{eq:rhoA_spectral}) is that the infinite-time average becomes a grouping problem over energy differences.  The only identity one needs is
\be
\lim_{T\to\infty}\frac1T\int_0^T\!\dd t\, e^{-i\Omega t}
=
\begin{cases}
1, & \Omega=0,\\
0, & \Omega\ne0 .
\end{cases}
\label{eq:time_avg_phase}
\ee
Since $\Tr\rho_A^2(t)$ is quadratic in $\rho_A(t)$, it contains phases of the form $\exp\{-i[(E_m-E_n)+(E_p-E_q)]t\}$.  Eq.~(\ref{eq:time_avg_phase}) therefore keeps precisely the terms satisfying $E_m-E_n=E_q-E_p$.  Thus the time average is sensitive not only to exact level degeneracies, but also to repeated energy gaps in an otherwise nondegenerate spectrum.

The calculation of entangling power can then be performed using the eigenvalue-grouping algorithm used in Ref.~\cite{Low:2026ep}.  First, diagonalize $H$ and arrange each eigenvector as the coefficient matrix $C_m$ in Eq.~(\ref{eq:Cm_def}).  Second, for the input $|\chi\>$, compute the energy-basis weights $c_m=\<E_m|\chi\>$.  Third, collect all ordered pairs $(m,n)$ with the same energy difference $\Omega=E_m-E_n$ into a group
\be
g_\Omega\equiv\{(m,n):E_m-E_n=\Omega\}.
\ee
Finally, for each group form a matrix $R_\chi(\Omega)$ by summing the contributions from all pairs in $g_\Omega$.  The time-averaged value of $\Tr\rho_A^2$ for the fixed input can then be evaluated as
\begin{align}
\left\<\Tr\rho_A^2\right\>_t
&=\sum_\Omega \left\|R_\chi(\Omega)\right\|_F^2,\nonumber\\
R_\chi(\Omega)
&=\sum_{\substack{m,n\\ E_m-E_n=\Omega}}
c_m c_n^*\,M_{mn}.
\label{eq:Rchi}
\end{align}
Here $\|\cdot\|_F$ denotes the Frobenius norm,
$\|X\|_F^2\equiv\Tr(X^\dagger X)$.  This expression is exact after grouping equal energy differences.  The first line follows because the time-average condition pairs the $\Omega$ and $-\Omega$ groups, while $R_\chi(-\Omega)=R_\chi(\Omega)^\dagger$.  It also treats exact degeneracies through projectors onto common energy-difference groups rather than through arbitrary choices of eigenvectors inside a degenerate subspace.

Eq.~(\ref{eq:Rchi}) makes explicit the two ingredients that determine the time-averaged entangling power.  The first ingredient is the eigenvalue spectrum.  The spectrum controls the groups $g_\Omega$: exact degeneracies enlarge the zero-frequency group, while repeated gaps enlarge groups at nonzero $\Omega$.  Enlarging a group means that more ordered pairs contribute to the same matrix $R_\chi(\Omega)$ before the Frobenius norm is taken.  The squared norm $\|R_\chi(\Omega)\|_F^2$ then contains cross terms between distinct pairs in the same group.  This is the spectral mechanism behind the minima in the entangling power in high-energy scattering~\cite{Low:2021ufv} and in many-body systems~\cite{Low:2026ep}: larger energy-difference groups tend to increase $\langle\Tr\rho_A^2\rangle_t$ and hence decrease the entangling power.

The second ingredient is the eigenvector structure.  Once the spectrum has decided which ordered pairs belong to a given group $g_\Omega$, the size of that group's contribution is determined by the subsystem-$A$ matrices $M_{mn}$ defined in Eq.~(\ref{eq:Mmn_def}), together with the input weights $c_m c_n^*$.  The matrices $M_{mn}$ are the subsystem-$A$ remnants of the eigenstate outer products after tracing out subsystem $B$.  Thus the entangling power can change even when the groups $g_\Omega$ are unchanged, provided the eigenvectors are reorganized.  We will see that the AQRM peaks presented below are of this second kind.  At generic fixed $(g,\Del)$, the integer-bias condition does not enlarge the zero-frequency group, but it reshapes the eigenvectors to produce sharp peaks.

When applying the above algorithm to a subsystem such as the harmonic oscillator, there is a subtlety associated with the fact that the Hilbert space is infinite-dimensional and has no normalizable Haar measure.  A Fock-space truncation to the first $\Ntr$ oscillator states, ${\cal H}_{B,\Ntr}=\mathrm{span}\{|0\>,\ldots,|\Ntr-1\>\}$, makes the integral finite, and at any fixed $\Ntr$ the truncated bosonic Haar average is a well-defined finite-dimensional quantity.
It is instructive to compute the average occupation number of the oscillator input produced by this truncated ensemble.  If $\dd\mu_{\rm H,\Ntr}(\phi)$ denotes the normalized Haar measure on ${\cal H}_{B,\Ntr}$, define
\begin{align}
\bar\rho_{B,\Ntr}
&\equiv \int\dd\mu_{\rm H,\Ntr}(\phi)\,|\phi\>\<\phi|
=\frac{\Id_{\Ntr}}{\Ntr} \ ,
\label{eq:haar_one_copy_state}
\end{align}
then the Haar-averaged oscillator occupation $\bar n_{\rm H}$ is
\begin{align}
\bar n_{\rm H}&\equiv \int\dd\mu_{\rm H,\Ntr}(\phi)\,
\<\phi|\adag a|\phi\> \nonumber\\
&=\Tr_{\Ntr}\!\left(\adag a\,\bar\rho_{B,\Ntr}\right)
 \nonumber\\
&=\frac{1}{\Ntr}\Tr_{\Ntr}(\adag a)
 \nonumber\\
&=\frac{1}{\Ntr}\sum_{n=0}^{\Ntr-1}n
=\frac{\Ntr-1}{2}.
\label{eq:haar_energy}
\end{align}
where $\Tr_{\Ntr}$ is the trace over ${\cal H}_{B,\Ntr}$.  In the last line we used the number-operator eigenvalues $\adag a|n\>=n|n\>$ within the truncated basis.  Eq.~(\ref{eq:haar_energy}) gives the important calibration between the cutoff and the oscillator scale being sampled: a finite-Haar diagnostic at large cutoff $\Ntr$ has mean input occupation $\bar n_{\rm H}\approx \Ntr/2$.  We will see below that the relevant scale for resolving the integer-bias peaks in the time-averaged entangling power is the mean input occupation, not the cutoff by itself.

Alternatively, one can replace the bosonic Haar average by a finite-energy input ensemble whose mean occupation is fixed independently of the Fock cutoff $\Ntr$.  A natural choice is an ensemble of coherent states, defined by
\begin{align}
|\alpha\>
&=e^{-|\alpha|^2/2}\sum_{n=0}^{\infty}
\frac{\alpha^n}{\sqrt{n!}}|n\>,\nonumber\\
a|\alpha\>&=\alpha|\alpha\>,\qquad
\<\alpha|\adag a|\alpha\>=|\alpha|^2 .
\label{eq:coherent_state_def}
\end{align}
Here $\alpha$ is the coherent-state amplitude, i.e.~the eigenvalue of the annihilation operator, and $|\alpha|^2$ is the mean oscillator occupation of the input.  Since coherent states form an overcomplete family rather than an orthonormal basis, specifying such an ensemble requires a normalized weight.  We denote this weight by $\mu_{\nbar}(\alpha)$ and require
\be
\int\!\dd^2\alpha\,\mu_{\nbar}(\alpha)=1,\qquad
\int\!\dd^2\alpha\,\mu_{\nbar}(\alpha)|\alpha|^2=\nbar .
\label{eq:coherent_weight_norm}
\ee
We take the bosonic inputs to be coherent states $|\alpha\>$ drawn from $\mu_{\nbar}(\alpha)$.  Coherent states are the most classical pure states of a harmonic oscillator and are the natural drive states in cavity and circuit QED, so this choice is physically well motivated.

Coherent-state sampling itself is not new: spin-coherent initial states have been used in quantum-chaos diagnostics, and coherent or Gaussian states enter continuous-variable notions of entangling capability and entanglement potential~\cite{WangChaos2004,Wolf2003,Asboth2005}.  Here the role of the coherent ensemble is more specific: it provides a finite-energy replacement for the oscillator Haar input measure in a Zanardi–Zalka–Faoro entangling-power average.

With this normalized weight, define
\be
\mathrm{ep}_{\rm coh}(U;\nbar)=
2\int\!\dd\mu_{\rm H}(\psi)\int\!\dd^2\alpha\,\mu_{\nbar}(\alpha)
\left[1-\Tr \rho_A^2\right],
\label{eq:ep_coh}
\ee
with $\rho_A$ still given by Eq.~(\ref{eq:rhoA_def}), now for the pure input $|\psi\>\otimes|\alpha\>$. The factor of $2$ is again $d_A/(d_A-1)$ for a qubit.  The infinite-time average is defined similarly,
\be
\epc(\nbar)\equiv
\lim_{T\to\infty}\frac{1}{T}\int_0^T\!\dd t\,
\mathrm{ep}_{\rm coh}\!\left(e^{-iHt};\nbar\right).
\label{eq:ep_coh_time}
\ee
In the numerical results below, we use the fixed-radius ensemble. Writing $\alpha=\sqrt{\nbar}e^{i\theta}$, this ensemble fixes $|\alpha|^2=\nbar$ and samples the phase $\theta$ uniformly. Explicitly, the normalized weight is the Delta-function measure
\be
\mu_{\nbar}^{\rm fr}(\alpha)
=\frac{1}{\pi}\,\delta\!\left(|\alpha|^2-\nbar\right).
\label{eq:fixed_radius_weight}
\ee
Then, for any integrand $F(\alpha)$,
\be
\int\!\dd^2\alpha\,\mu_{\nbar}^{\rm fr}(\alpha)F(\alpha)
=
\frac{1}{2\pi}\int_0^{2\pi}\!\dd\theta\,
F\!\left(\sqrt{\nbar}e^{i\theta}\right).
\label{eq:fixed_radius_mu}
\ee
A Gaussian coherent ensemble can also be defined; Appendix~\ref{app:coherent}
compares it with fixed-radius sampling at the same mean occupation.

In the numerical implementation, we evaluate the phase average in Eq.~(\ref{eq:fixed_radius_mu}) on a uniform grid; convergence with respect to the grid size was checked.  Moreover, a coherent state is a superposition over all Fock states, with Fock-space weight centered near $\nbar$ and width of order $\sqrt{\nbar}$.  Thus one still needs a truncated Fock space to represent the coherent state in numerics, although the cutoff $\Ntr$ only controls how much of this Fock-space weight is retained.  In the fixed-radius scans shown below, we choose the cutoff so that
\be
\frac{\Ntr-\nbar}{\sqrt{\nbar}}\sim \mathcal{O}(5\text{--}10),
\label{eq:coherent_cutoff_margin}
\ee
namely the truncation lies several coherent-state widths above the center of the Fock-space distribution.  The physical input scale is therefore set by $\nbar$, while $\Ntr$ acts only as a numerical cutoff.

Each numerical scan reported below required at most approximately one CPU core-hour.

\section{AQRM Integer-Bias Peaks}
\label{sec:aqrm_peaks}

Having defined the two input ensembles and the time-averaged diagnostic, we now use them to test the hidden-symmetry condition directly in the AQRM.  We begin by recalling the relevant coupling regimes, and then use numerical bias scans to identify the common integer-bias structure.  Finally, we explain the locations of the peaks using a displaced-doublet picture.

The light--matter coupling $g$ organizes the AQRM into several physically distinct regimes:
\begin{itemize}

\item The weak-coupling regime: $g/\om\ll 1$, where the counter-rotating terms in Eq.~(\ref{eq:rotating_counterrotating}) are perturbative and the dynamics remains close to the JC limit.
\item The ultrastrong-coupling regime: $g/\om\gtrsim 0.1$, where the counter-rotating terms are no longer removable by the RWA and must be retained in the full Hamiltonian.
\item The deep-strong-coupling regime: $g/\om\gtrsim 1$, where $\sx$ and $(a+a^{\dagger})$ become strongly correlated in the Hamiltonian eigenstates, which gives rise to a different perturbative regime.

\end{itemize}
The AQRM hidden symmetry is not tied to any one of these regimes: according to the calculation of Ref.~\cite{Mangazeev:2020yjs} presented in Sec.~\ref{sec:qrm_family}, for every integer bias $\ep/\om=k$, there exists a parameter-dependent operator $J_k$ commuting with the Hamiltonian.  It is therefore natural to ask whether the time-averaged entangling power sees a common integer-bias structure across these qualitatively different coupling regimes.

\begin{figure*}[t]
\centering
\includegraphics[width=\textwidth]{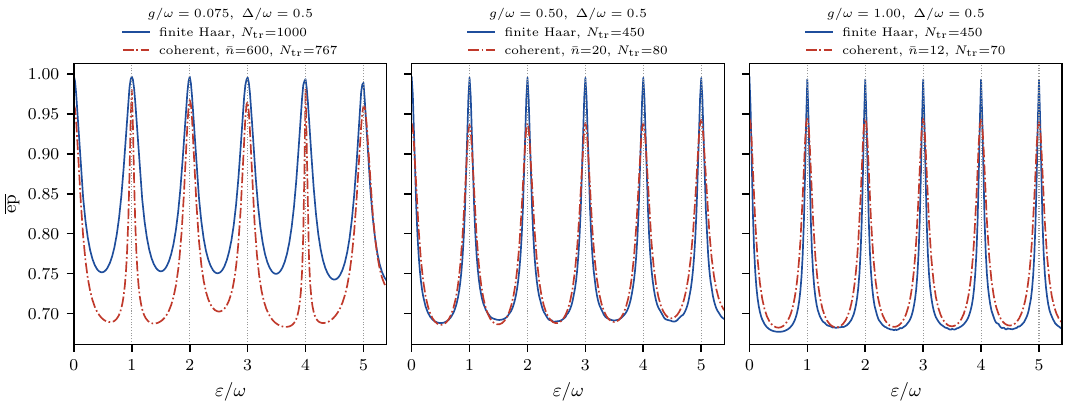}
\caption{Time-averaged entangling power $\epb$ versus bias $\ep/\om$ for the AQRM at $\Del/\om=0.5$.  The three panels correspond to three representative values of the coupling strength: $g/\om=0.075$, $0.50$, and $1.00$, spanning weak coupling, ultrastrong coupling, and deep-strong coupling.  Solid blue curves use a finite-Haar oscillator input at cutoff $\Ntr$, while dash-dot red curves use a fixed-radius coherent input at mean occupation $\nbar$.   The maxima in both ensembles occur at the integer biases $\ep/\om\in\mathbb{Z}$; the vertical dotted lines mark the integer values shown in the plotted range.}
\label{fig:sec4}
\end{figure*}

Figure~\ref{fig:sec4} gives the central numerical result.  We plot the time-averaged entangling power as a function of $\ep/\om$ at fixed $\Del/\om=0.5$, for three values of the coupling strength, each representing a different coupling regime: $g/\om=0.075$, $0.50$, and $1.00$.  These values probe, respectively, the weak-coupling regime, the ultrastrong-coupling regime, and the deep-strong-coupling regime.  In each panel, we compare two oscillator input measures: a finite-Haar average over a truncated Fock space, and the fixed-radius coherent-state ensemble introduced in Sec.~\ref{sec:ep_coherent}.  The two diagnostics weight the oscillator Hilbert space differently, and therefore they need not agree pointwise.  Nevertheless, in every case, the maxima occur at the same integer-bias points, $\ep/\om\in\mathbb{Z}$. The scans are performed at fixed $(g/\om,\Del/\om)$ and are not tuned to special Juddian constraint-polynomial loci.  Thus the integer-bias response is a robust feature of the AQRM dynamics at the hidden-symmetry values.

Several comments are in order. First, the agreement between the two diagnostics is a statement about peak location, not about the detailed height of the curves.  In the coherent ensemble, the physical input scale is the mean occupation $\nbar$, while $\Ntr$ only controls how much of the coherent-state Fock support is retained.  In the finite-Haar ensemble, by contrast, the cutoff also sets the typical oscillator occupation sampled by the input ensemble, $\bar n_{\rm H}=(\Ntr-1)/2$.  Changing $\Ntr$ therefore changes the diagnostic itself.  The fact that both diagnostics nevertheless locate the same integer-bias maxima shows that the peak positions are controlled by the AQRM eigenstructure rather than by the detailed choice of input measure.

Second, the mean occupation needed to resolve the first five peaks depends strongly on the coupling strength. At small $g$, the displacement of the two oscillator ladders is small, and resolving the relevant displaced doublets requires support at much larger occupation.  This is why the weak-coupling panel uses substantially larger coherent and finite-Haar supports than the $g=0.50$ and $g=1.00$ panels.  At larger coupling strength, the same five integer-bias peaks are already visible with modest coherent mean occupation.  This behavior anticipates the displaced-doublet explanation below: the integer locations follow from a rung-independent diagonal energy difference, while the required occupation support is controlled by the displaced-oscillator overlaps.

We now explain the origin of the integer locations at $\ep=k\om$, using the displaced-oscillator picture familiar from generalized rotating-frame treatments of the Rabi model~\cite{Irish:2007,AshhabNori:2010,Ashhab:2019xft}. The useful starting point is the $\sx$ eigenbasis,
\be
\sx|{\rightarrow}\>=|{\rightarrow}\>,\qquad
\sx|{\leftarrow}\>=-|{\leftarrow}\>.
\label{eq:sx_basis}
\ee
In this basis the bias and light--matter coupling are diagonal in the qubit label, while the qubit-splitting term $\Del\sz/2$ mixes the two qubit sectors.  We therefore first set aside the $\Del\sz/2$ term and study
\be
H_0=
\om\left(\adag a+\frac12\right)
+\frac{\ep}{2}\sx
+g\,\sx(a+\adag) \ ,
\label{eq:H0_displaced}
\ee
which is diagonal in the $\{|{\rightarrow}\>,|{\leftarrow}\>\}$ basis.
For a fixed $\sx$ eigenvalue, $H_0$ is a shifted harmonic oscillator.  In the $|{\rightarrow}\>$ sector one has
\be
H_{0,{\rightarrow}}
=
\om\left[
\left(\adag+\frac{g}{\om}\right)
\left(a+\frac{g}{\om}\right)
+\frac12
\right]
-\frac{g^2}{\om}
+\frac{\ep}{2},
\label{eq:H0_right}
\ee
while in the $|{\leftarrow}\>$ sector
\be
H_{0,{\leftarrow}}
=
\om\left[
\left(\adag-\frac{g}{\om}\right)
\left(a-\frac{g}{\om}\right)
+\frac12
\right]
-\frac{g^2}{\om}
-\frac{\ep}{2}.
\label{eq:H0_left}
\ee
Thus the unperturbed problem consists of two harmonic-oscillator ladders, displaced in opposite directions and shifted in energy by the bias.  The corresponding eigenstates at $\Delta=0$ may be written as
\begin{align}
|{\rightarrow},j\>_d&=|{\rightarrow}\>D(-g/\om)|j\>,\nonumber\\
|{\leftarrow},j\>_d&=|{\leftarrow}\>D(g/\om)|j\>,
\label{eq:displaced_states}
\end{align}
where $j=0,1,2,\ldots$ labels the oscillator level in the displaced ladder and
\be
D(\alpha)=\exp(\alpha\adag-\alpha^* a)
\label{eq:displacement_operator}
\ee
is the displacement operator.  It translates the oscillator operators by a c-number,
\be
D^\dagger(\alpha)aD(\alpha)=a+\alpha .
\label{eq:displacement_action}
\ee
For the real displacements appearing here, this is a shift of the oscillator wavefunction in position.  In particular,
\be
\begin{aligned}
\left(a+\frac{g}{\om}\right)D(-g/\om)|j\>
\ &=D(-g/\om)a|j\>,\\
\left(a-\frac{g}{\om}\right)D(g/\om)|j\>
\ &=D(g/\om)a|j\>.
\end{aligned}
\label{eq:displaced_number_identity}
\ee
This is why the two states in Eq.~(\ref{eq:displaced_states}) provide a convenient basis for the finite-$\Del$ problem.  In the $\sx$ basis, the part $H_0$ is diagonal in the displaced states, while $\sz$ flips the two $\sx$ labels:
\be
\begin{aligned}
\<{\rightarrow}|\sz|{\rightarrow}\>
&=\<{\leftarrow}|\sz|{\leftarrow}\>=0,\\
\<{\rightarrow}|\sz|{\leftarrow}\>
&=\<{\leftarrow}|\sz|{\rightarrow}\>=1 .
\end{aligned}
\label{eq:sz_sx_matrix_elements}
\ee
Thus the diagonal entries of the two-state problem come from $H_0$, whereas the finite-$\Del$ term supplies the off-diagonal entry.  The corresponding diagonal energies are
\begin{align}
E_{{\rightarrow},j}^{(0)}
&=\om\left(j+\frac12\right)-\frac{g^2}{\om}+\frac{\ep}{2},\nonumber\\
E_{{\leftarrow},j}^{(0)}
&=\om\left(j+\frac12\right)-\frac{g^2}{\om}-\frac{\ep}{2}.
\label{eq:displaced_energies}
\end{align}

We now project the finite-$\Del$ Hamiltonian onto the displaced pair
\be
|{\rightarrow},j\>_d,\qquad |{\leftarrow},j+k\>_d .
\label{eq:jk_pair}
\ee
The difference of the two diagonal entries in this two-state subspace is
\be
\eta_k
\equiv
E_{{\rightarrow},j}^{(0)}-E_{{\leftarrow},j+k}^{(0)}
=\ep-k\om .
\label{eq:energy_difference}
\ee
The off-diagonal entry is instead generated by the finite-$\Del$ term:
\be
{}_{d}\<{\rightarrow},j|
\frac{\Del}{2}\sz
|{\leftarrow},j+k\>_{d}
=\frac{\Del}{2}\<j|D(2g/\om)|j+k\>.
\label{eq:offdiag_overlap}
\ee
We define the oscillator overlap appearing in this matrix element as
\be
s_{jk}=\<j|D(2g/\om)|j+k\>,
\label{eq:displaced_overlap}
\ee
so that the off-diagonal entry is $\Del s_{jk}/2$.  When the pair in Eq.~(\ref{eq:jk_pair}) is sufficiently isolated from neighboring displaced states, it can be treated as an effective two-level system.\footnote{More explicitly, the matrix elements coupling this pair to other displaced states must be small compared with the corresponding energy separations.}  Under this condition, combining the diagonal energy difference and the off-diagonal entry, and subtracting the average energy, gives the effective Hamiltonian
\be
H_{\rm eff}^{(j,k)}=\frac12
\begin{pmatrix}
\eta_k & \Del s_{jk}\\
\Del s_{jk} & -\eta_k
\end{pmatrix} .
\label{eq:Heff}
\ee
Diagonalizing Eq.~(\ref{eq:Heff}) gives the mixing angle
\be
\sin^2 2\theta_{jk}=
\frac{\Del^2 |s_{jk}|^2}{\eta_k^2+\Del^2 |s_{jk}|^2} ,
\label{eq:mixing_angle}
\ee
For nonzero overlap $s_{jk}$, Eq.~(\ref{eq:mixing_angle}) is maximized at $\sin^2 2\theta_{jk}=1$ when $\eta_k=0$.  Using Eq.~(\ref{eq:energy_difference}), this gives
\be
\ep=k\om .
\label{eq:integer_location}
\ee
The important point is that this condition is independent of $j$, so the same integer bias maximizes the mixing for the entire family of doublets.  This is the location mechanism for the peaks shown in Fig.~\ref{fig:sec4}.  In the next section we will see that the size and visibility of a given peak depend on the overlaps $s_{jk}$ and on whether the input ensemble has appreciable weight on the relevant oscillator levels.

\section{Why the Integer-bias Feature Is a Peak}
\label{sec:sign}

The displaced-doublet analysis in the previous section explains why the special bias points are located at
$\ep=k\om$.  This location argument does not, by itself, determine the sign of the feature in
entangling power.  This is an important distinction.  In several examples, including scattering and spin chains, enhanced symmetry is associated with a dip in entangling power~\cite{Low:2021ufv,Liu:2022grf,Carena:2023vjc,Low:2026ep}.\footnote{In one instance, Ref.~\cite{Carena:2025wyh}, maximizing entanglement does lead to enhanced symmetries.}  The AQRM integer-bias feature in Fig.~\ref{fig:sec4} is instead a peak.  In this section we explain this sign difference by separating two effects in the time-averaged purity.  A symmetry can act spectrally, by producing exact degeneracies or repeated gaps, or it can act through the eigenvectors that enter the reduced density matrices.  For the AQRM bias scans considered here, the peak is tied to the latter effect.

The basic point is simple.  In the present normalization,
$\epb=2(1-\langle\Tr\rho_A^2\rangle_t)$, with the final
average also taken over the chosen product-input ensemble.  Thus a peak in
entangling power is a dip in the time-averaged value of $\Tr\rho_A^2$.  From
Eq.~(\ref{eq:Rchi}), this quantity is built by grouping ordered eigenstate pairs
according to their energy difference $\Omega=E_m-E_n$.  A conventional symmetry tends to produce exact degeneracies or repeated gaps and thereby enlarge these groups, placing more ordered pairs in the same
$\Omega$ group.  Contributions from these pairs are summed inside
$R_\chi(\Omega)$ before the Frobenius norm $\|\cdot\|_F^2$ is taken. The resulting
 norm  tends to  raise
$\langle\Tr\rho_A^2\rangle_t$, producing a dip in entangling power.  This spectral route drives the spin-chain dips of Ref.~\cite{Low:2026ep}; we will see in Sec.~\ref{sec:jc} that it is not the only route to a dip. In the fixed-parameter bias scans shown in Fig.~\ref{fig:sec4}, however, the integer-bias condition does not generically produce exact degeneracies or repeated gaps: the scans vary $\ep/\om$ at fixed $(g/\om,\Del/\om)$ and are not tuned to Juddian constraint-polynomial loci.  The peak must therefore come from the second ingredient in Eq.~(\ref{eq:Rchi}): the eigenvectors.

We first describe the eigenvector channel in the finite-Haar diagnostic, where
it has a compact invariant form.  For each energy eigenstate $|E_m\>$, let
$C_m$ be the coefficient matrix in Eq.~(\ref{eq:Cm_def}).  We use the same
product-basis notation as in Eq.~(\ref{eq:Cm_def}): $|a,b\>\equiv
|a\>_A\otimes |b\>_B$, with $a=1,2$ labeling the qubit and
$b=0,\ldots,\Ntr-1$ labeling the truncated oscillator basis.  Then
\be
|E_m\>=\sum_{a=1}^{2}\sum_{b=0}^{\Ntr-1}(C_m)_{ab}|a,b\>,\qquad
\Pi_m\equiv |E_m\>\<E_m| .
\label{eq:eigen_projector}
\ee
 The density matrix of the qubit subsystem  is obtained by tracing over
the oscillator index,
\begin{align}
\rho_A^{(m)}
&\equiv\Tr_B\Pi_m \nonumber\\
&=\sum_{a,a'=1}^{2}\sum_{b=0}^{\Ntr-1}
(C_m)_{ab}(C_m)^*_{a'b}|a\>\<a'| \nonumber\\
&=C_m C_m^\dagger .
\label{eq:rhoA_eigenstate}
\end{align}
In the notation of Eq.~(\ref{eq:Mmn_def}), this is the diagonal object
$\rho_A^{(m)}=M_{mm}$.
The resulting reduced density matrix is $2\times2$ and can be parameterized by a Bloch vector,
\be
\rho_A^{(m)}=\frac12\left(\Id+\rvec_m\cdot\bm\sigma\right).
\label{eq:rho_bloch}
\ee
In the finite-Haar formulation, the Haar average of $\Tr\rho_A^2$ is expressed
in terms of two swap invariants, conventionally denoted $I_0$ and $I_1$; their
frequency-group form is reviewed in Appendix~\ref{app:swap_formalism}.  Here we
only need the $\Omega=0$ diagonal-pair contribution to
Eq.~(\ref{eq:I0_group_app}), for which $M_{mm}=\rho_A^{(m)}$.
Away from the integer bias, the zero-frequency group contains only the
diagonal pairs $(m,m)$ generically.  The diagonal part of the finite-Haar $I_0$ is given by
\be
I_0^{\rm diag}=\sum_{mn}\left(\Tr\big[\rho_A^{(m)}\rho_A^{(n)}\big]\right)^2 .
\label{eq:I0_diag_def}
\ee
Using
$\Tr[\rho_A^{(m)}\rho_A^{(n)}]=\tfrac12(1+\rvec_m\cdot\rvec_n)$, and summing
over the $d=2\Ntr$ eigenstates of the truncated qubit--oscillator Hilbert
space, Eq.~(\ref{eq:I0_diag_def}) becomes
\be
I_0^{\rm diag}=\frac14\left[d^2+2|\mathbf R|^2+\|Q\|_F^2\right],
\label{eq:I0_diag}
\ee
where the Bloch dipole and quadrupole are
\be
\mathbf R=\sum_m\rvec_m,\qquad
Q_{ab}=\sum_m r_m^a r_m^b .
\label{eq:RQ}
\ee
The first moment $\mathbf R$ vanishes by completeness.  Indeed,
$\sum_m\rho_A^{(m)}=\Tr_B\Id=\Ntr\,\Id$, and the traceless part gives
$\mathbf R=0$.  Thus, in this diagonal finite-Haar channel, all nontrivial
eigenvector dependence sits in the Bloch quadrupole $\|Q\|_F^2$.  This formula
should be viewed as a diagnostic of the dominant moving channel, not as a
closed expression for the full observable; the full time-averaged entangling
power also contains the remaining swap term and the nonzero-frequency groups.

The displaced doublets of Sec.~\ref{sec:aqrm_peaks} give the microscopic
origin of the suppression of this channel.  Consider one pair
$|{\rightarrow},j\>_d$ and $|{\leftarrow},j+k\>_d$.  An eigenstate of the
effective Hamiltonian in Eq.~(\ref{eq:Heff}) has the form
\be
|\psi_{jk}\>=\cos\theta_{jk}\,|{\rightarrow},j\>_d+\sin\theta_{jk}\,|{\leftarrow},j+k\>_d ,
\label{eq:doublet_eigenstate}
\ee
where the mixing angle is given in Eq.~(\ref{eq:mixing_angle}).  As above, one
must distinguish the rank-one density matrix
$|\psi_{jk}\>\<\psi_{jk}|$ on the full qubit--oscillator Hilbert space from the
reduced density matrix of the qubit.  The latter can be mixed, because the
oscillator trace discards which oscillator state is attached to each qubit
component.  It is useful to write explicitly what this trace does.  Using
Eq.~(\ref{eq:displaced_states}), the two qubit components in
Eq.~(\ref{eq:doublet_eigenstate}) are attached to the oscillator states
$D(-g/\om)|j\>$ and $D(g/\om)|j+k\>$.  The diagonal terms of
$|\psi_{jk}\>\<\psi_{jk}|$ therefore give the two $\sx$-sector populations,
while the cross terms retain only the inner product of the two oscillator
states.  Thus
\begin{align}
\rho_{A,jk}
&=\cos^2\theta_{jk}|{\rightarrow}\>\<{\rightarrow}|
 +\sin^2\theta_{jk}|{\leftarrow}\>\<{\leftarrow}| \nonumber\\
&\quad
 +\cos\theta_{jk}\sin\theta_{jk}
 \left(s_{jk}^*|{\rightarrow}\>\<{\leftarrow}|
 +s_{jk}|{\leftarrow}\>\<{\rightarrow}|\right),
\label{eq:rho_doublet_reduced}
\end{align}
where $s_{jk}=\<j|D(2g/\om)|j+k\>$ is the overlap defined in
Eq.~(\ref{eq:displaced_overlap}).  The first line of
Eq.~(\ref{eq:rho_doublet_reduced}) gives a Bloch component
$\cos^2\theta_{jk}-\sin^2\theta_{jk}=\cos2\theta_{jk}$ along the $\sx$
direction.  The second line is the off-diagonal term between the two $\sx$ sectors; it
is reduced by $s_{jk}$ because the oscillator states attached to
$|{\rightarrow}\>$ and $|{\leftarrow}\>$ are not the same.  In a phase
convention where $s_{jk}$ is real, the corresponding Bloch vector is, up to an
overall sign,
\be
\rvec_{jk}=\left(\cos 2\theta_{jk},\,0,\,s_{jk}\sin 2\theta_{jk}\right),
\label{eq:rvec_doublet}
\ee
and hence, using Eq.~(\ref{eq:mixing_angle}) in the second line,
\begin{align}
|\rvec_{jk}|^2
&=
1-\left(1-|s_{jk}|^2\right)\sin^2 2\theta_{jk} \nonumber\\
&=
1-\left(1-|s_{jk}|^2\right)
\frac{\Del^2 |s_{jk}|^2}{\eta_k^2+\Del^2 |s_{jk}|^2}.
\label{eq:rvec_floor}
\end{align}
This equation is the useful physical picture.  The second line shows directly
that when $|\eta_k| \gg |\Del s_{jk}|$, 
$|\rvec_{jk}|^2\to 1$.  At $\ep=k\om$,
Eq.~(\ref{eq:energy_difference}) gives $\eta_k=0$.  For a doublet with nonzero overlap $s_{jk}$, 
\be
|\rvec_{jk}|^2=|s_{jk}|^2 .
\label{eq:overlap_floor}
\ee

\begin{figure*}[t]
\centering
\includegraphics[width=0.92\textwidth]{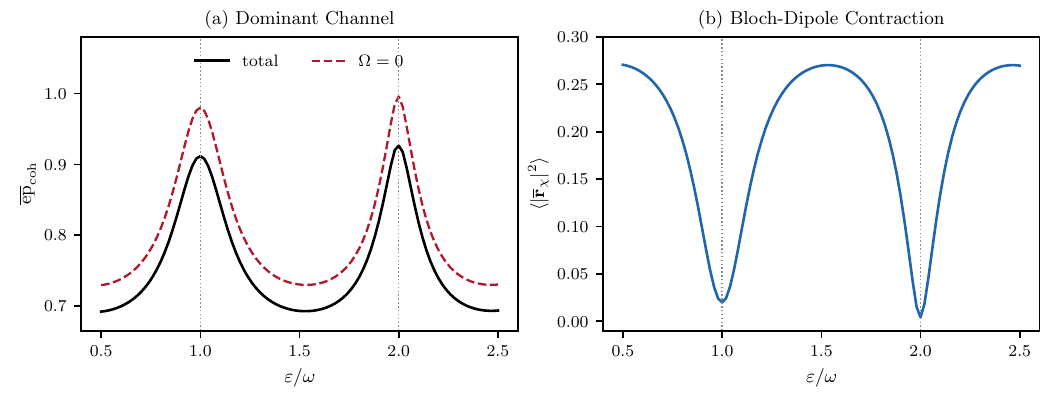}
\caption{Coherent-channel decomposition for $g/\om=0.30$, $\Del/\om=0.5$, $\nbar=20$, and $\Ntr=60$.  Panel (a) shows the total $\epc$ and the $\Omega=0$ contribution $2(1-\<P_0\>)$; panel (b) shows the corresponding Bloch-dipole diagnostic $\<|\avg{\rvec}_\chi|^2\>$.  The comparison shows that the integer-bias peaks are already present in the zero-frequency channel.  The simultaneous dips in $\<|\avg{\rvec}_\chi|^2\>$ show that these peaks are driven by contraction of the population-weighted Bloch dipole, i.e. by eigenvector reorganization rather than by additional degeneracies.}
\label{fig:coherent_channel}
\end{figure*}

The overlap $s_{jk}$ is the inner product of two normalized oscillator states,
so the Cauchy--Schwarz inequality gives $|s_{jk}|\le 1$.  Equality would require
$D(2g/\om)|j+k\>$ to be the same vector as $|j\>$ up to an overall phase.  For
the displaced pairs of interest this is not true in general: the two oscillator
components differ both by the relative displacement and by the oscillator
level.  Thus $|s_{jk}|<1$ generically, and the integer-bias point shortens the
reduced-qubit Bloch vector for this doublet, signaling stronger
qubit--oscillator entanglement.

This shortening of many Bloch vectors at once is the origin of the peak.  The
quantity $\|Q\|_F^2$ in Eq.~(\ref{eq:I0_diag}) is large when the eigenstate
	Bloch vectors are long and aligned in Bloch space.  The integer
condition shortens the vectors associated with the displaced doublets and
therefore lowers the diagonal contribution to $\langle\Tr\rho_A^2\rangle_t$.
Since the entangling power is $2(1-\langle\Tr\rho_A^2\rangle_t)$, lowering this
channel raises the entangling power.  It is worth emphasizing the collective
nature of the effect: the energy difference
$\eta_k=\ep-k\om$ is independent of the oscillator level $j$, so a whole family of
doublets reaches maximal mixing at the same bias.  The observed peak is not
	the response of a single avoided crossing, but the collective accumulation of
many eigenvector reorganizations at the same integer value of $\ep/\om$.

The finite-Haar discussion above is controlled by the unweighted Bloch-vector quadrupole $Q$.  For a coherent-state input, the diagonal channel is instead controlled by a population-weighted average of the same eigenstate Bloch vectors.  To see this, fix one product input $|\chi\>$.
At generic bias, the $\Omega=0$ group in Eq.~(\ref{eq:Rchi}) again consists only of the diagonal pairs. Therefore setting $\Omega=0$ in Eq.~(\ref{eq:Rchi}) gives
\begin{align}
R_\chi(0)
&=\sum_{\substack{m,n\\E_m=E_n}}
c_m c_n^*\,M_{mn} \nonumber\\
&=\sum_m |c_m|^2 M_{mm}
=\sum_m p_m\,\rho_A^{(m)} .
\label{eq:Rchi0}
\end{align}
In the second line we used the fact that, at generic bias, the relevant part
of the spectrum is nondegenerate, so the zero-frequency group contains only
the diagonal pairs $(m,m)$.  We also used
$p_m=|c_m|^2=|\<E_m|\chi\>|^2$ and
$M_{mm}=\rho_A^{(m)}$.  Equivalently,
the full diagonal ensemble for the evolved input is
\be
\bar\rho_\chi=\sum_m p_m |E_m\>\<E_m| .
\label{eq:diag_ensemble_chi}
\ee
Tracing out the oscillator gives
\be
\Tr_B\bar\rho_\chi
=\sum_m p_m M_{mm}
=\sum_m p_m\rho_A^{(m)}
=R_\chi(0) .
\label{eq:Rchi0_diag_ensemble}
\ee
Thus, in the generic case with no exact degeneracies, $R_\chi(0)$ is the
diagonal-ensemble reduced qubit state, which we denote by
$\bar\rho_A$.  Writing
$\bar\rho_A=\tfrac12(\Id+\avg{\rvec}_\chi\cdot\bm\sigma)$ gives
\be
\left\|R_\chi(0)\right\|_F^2=\Tr\bar\rho_A^2=\frac12\left(1+|\avg{\rvec}_\chi|^2\right),
\qquad
\avg{\rvec}_\chi=\sum_m p_m\rvec_m .
\label{eq:coh_dipole}
\ee
The coherent diagonal channel is therefore governed by a population-weighted
Bloch dipole $\avg{\rvec}_\chi$, rather than by the unweighted finite-Haar
quadrupole $Q$.  This difference is important technically, but it does not
change the physical mechanism.  At integer bias, the doublet Bloch vectors
shorten to the overlap-controlled value in Eq.~(\ref{eq:overlap_floor}).  For coherent inputs with support on
those doublets, the weighted dipole $\avg{\rvec}_\chi$ also contracts,
$\bar\rho_A$ moves toward $\Id/2$, and the zero-frequency
contribution to $\langle\Tr\rho_A^2\rangle_t$ decreases.  The coherent
entangling power therefore peaks for the same reason as the finite-Haar
diagnostic: the integer-bias hidden-symmetry point is where finite-$\Del$
mixing most efficiently entangles the qubit with the oscillator within the
relevant displaced doublets.

Figure~\ref{fig:coherent_channel} displays the two main ingredients of this
explanation for the coherent observable at $g/\om=0.30$, $\Del/\om=0.5$.  Panel (a)
compares the total $\epc$ with the contribution from the $\Omega=0$ group in
Eq.~(\ref{eq:Rchi}).  The difference between the two curves is exactly the sum
of the $\Omega\ne0$ groups, which is small and negative over this window.
Panel (b) plots $\<|\avg{\rvec}_\chi|^2\>$, showing directly that the
population-weighted Bloch dipole contracts at the integer-bias points.  By
Eq.~(\ref{eq:coh_dipole}), this contraction lowers the diagonal contribution to
$\langle\Tr\rho_A^2\rangle_t$.  Since the coherent entangling power is
$2(1-\langle\Tr\rho_A^2\rangle_t)$, the same contraction raises $\epc$.

The peaks in Fig.~\ref{fig:coherent_channel} are not coming from extra
eigenvalue degeneracies.  If two distinct eigenstates became exactly
degenerate, additional off-diagonal pairs $(m,n)$ with $m\ne n$ would enter the
$\Omega=0$ group in Eq.~(\ref{eq:Rchi}).  This does not occur in the scan shown:
the smallest spacing between distinct levels is $2.03\times10^{-3}\om$, and the
numerically constructed $\Omega=0$ group contains exactly $d=2\Ntr=120$ pairs,
one for each diagonal pair $(m,m)$ in the truncated qubit--oscillator Hilbert
space used in Fig.~\ref{fig:coherent_channel}.  Thus the integer-bias peak is
not a frequency-grouping effect.  It is an eigenvector effect, controlled by
the contraction of the population-weighted Bloch dipole.

\section{Jaynes--Cummings Point}
\label{sec:jc}

\begin{figure*}[t!]
\centering
\includegraphics[width=0.92\textwidth]{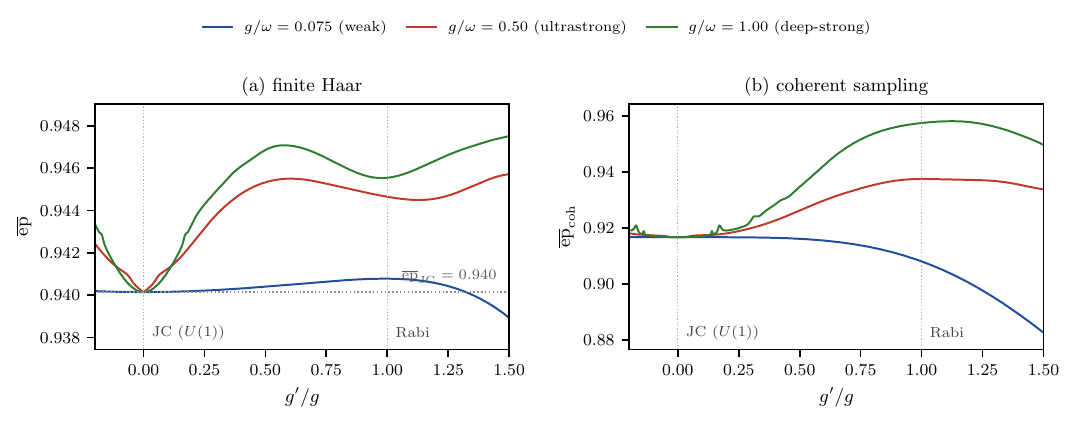}
\caption{Time-averaged entangling power across the generalized Rabi family at $\om=\Del=1$.  The three couplings are the same as in Fig.~\ref{fig:sec4}: $g/\om=0.075$, $0.50$, and $1.00$.  Panel (a) shows the finite-Haar diagnostic with $\Ntr=30$.    Panel (b) shows the fixed-radius coherent diagnostic with $\nbar=20$ and $\Ntr=80$.  The JC point $g'/g=0$ restores the $U(1)$  symmetry, while  $g'/g=1$ is the standard Rabi model with parity symmetry. }
\label{fig:jc}
\end{figure*}

We close the main analysis by considering the JC limit of the generalized Rabi
family as a control case.  In contrast to the AQRM hidden symmetry discussed
above, the JC symmetry is visible directly in the Hamiltonian: at $g'=0$ the
excitation number is conserved and the Hilbert space decomposes into fixed-charge
sectors.  This provides a useful comparison with the integer-bias peaks of
Sec.~\ref{sec:aqrm_peaks}, because the same diagnostic now responds with a dip
rather than a peak.

The purpose of this section is to explain both the sign and the size of this
response.  The JC dip is a weak feature whose contrast decreases with the input
scale, whereas the AQRM integer-bias peaks remain order-one effects.  We will
see that the difference reflects which part of the time-averaged purity is
dominant: the AQRM peaks are controlled by eigenvector contraction in the
diagonal channel, while the JC dip is controlled by the off-diagonal blocks
allowed by the fixed-charge sector structure.

We use the anisotropic Hamiltonian of Eq.~(\ref{eq:anisotropic_qrm}) at zero
bias, writing $g\equiv g_r$ and $g'\equiv g_{\rm cr}$, and scan the ratio
$g'/g$ at fixed $g$.  At the JC point $g'=0$ the excitation
number $\hat N=\adag a+\sigma_+\sigma_-$ is conserved, and the Hilbert space
decomposes into two-dimensional charge sectors spanned by
$\left|\uparrow,n\right\>$ and $\left|\downarrow,n+1\right\>$, together with
the one-dimensional sector containing $\left|\downarrow,0\right\>$.  This
sector structure follows from the conserved $\hat N$ alone and is therefore
present for any $g$, $\om$, and $\Del$; it is the relevant contrast with the
parameter-dependent hidden symmetry of Sec.~\ref{sec:qrm_family}.  We further
specialize to resonance, $\om=\Del$, where the two states in each sector are
degenerate before the coupling is turned on; for any nonzero $g$ the doublet
eigenvectors then take the $g$-independent equal-superposition form displayed
below in Eq.~(\ref{eq:jc_eigs}).  Off resonance the conserved charge survives;
in the finite-Haar regimes we examined, the JC point remains a local minimum,
but within each sector the eigenvectors acquire a $g$-dependent mixing angle,
so the value of $\epb$ at the JC point is no longer independent of the
coupling; the closed form derived below holds only at resonance.

At $\om=\Del$ and $g\ne0$, the spectrum is built from JC doublets
\be
\left|n,\pm\right\rangle
=\frac{1}{\sqrt2}\left(
\left|\uparrow,n\right\rangle
\pm
\left|\downarrow,n+1\right\rangle
\right),
\label{eq:jc_eigs}
\ee
for $n=0,\ldots,\Ntr-2$, with eigenvalues
$E_{n,\pm}=\om\left(n+1\right)\pm g\sqrt{n+1}$, together with two boundary
singlets, $\left|\downarrow,0\right\rangle$ and
$\left|\uparrow,\Ntr-1\right\rangle$.  The first is the physical JC ground
state, while the second is the upper edge state introduced by the finite Fock
truncation; both are kept explicitly in the finite-$\Ntr$ calculation below.
Each doublet eigenstate is an equal superposition of two states with different
qubit levels and different oscillator occupations.  After tracing out the
oscillator, the reduced qubit density matrix is $\rho_A=\Id/2$, so its Bloch vector
vanishes.  In this respect the JC point looks even more extreme than the AQRM
integer point: the doublet Bloch vectors do not merely shorten to an
overlap-controlled nonzero value, but disappear entirely.  At finite cutoff
only the two boundary singlets carry nonzero Bloch vectors, so the diagonal
channel of Sec.~\ref{sec:sign} exceeds its absolute floor of $1/2$ only
through an edge-state correction of order $\Ntr^{-2}$, as shown in
Appendix~\ref{app:jc}.  If this diagonal channel were the whole story, the JC
point would be a maximal peak.  We next consider the off-diagonal terms in the
same spectral expansion.

The relevant blocks are the matrices $M_{mn}$ of Eq.~(\ref{eq:Mmn_def}),
which appear in Eq.~(\ref{eq:rhoA_spectral}).  The diagonal block
$M_{mm}=\rho_A^{(m)}$ is the reduced qubit density matrix of eigenstate $m$ and its weight
$\|M_{mm}\|_F^2=\Tr[(\rho_A^{(m)})^2]$ is the purity.  The off-diagonal blocks
$M_{mn}$ with $m\ne n$ multiply the oscillating phase $e^{-i(E_m-E_n)t}$ in
Eq.~(\ref{eq:rhoA_spectral}).  Completeness of the energy eigenbasis,
equivalently $\sum_nC_n^\dagger C_n=d_A\Id_B$, gives
\be
\begin{aligned}
\sum_{n}\left\|M_{mn}\right\|_F^2
&=\Tr\left(C_m\left[\sum_nC_n^\dagger C_n\right]C_m^\dagger\right)\\
&=d_A\Tr\rho_A^{(m)}
=d_A=2,
\qquad\text{for every }m \ .
\end{aligned}
\label{eq:offdiag_sum_rule}
\ee
Separating the $n=m$ term gives
\be
\underbrace{\sum_{n\ne m}\left\|M_{mn}\right\|_F^2}_{\text{off-diagonal weight}}
=d_A-\Tr\big[(\rho_A^{(m)})^2\big] \ ,
\label{eq:offdiag_weight}
\ee
Thus reducing $\Tr[(\rho_A^{(m)})^2]$ increases the total squared weight of the
off-diagonal reduced blocks by the same amount.  The maximally
mixed JC doublets, with $\rho_A=\Id/2$, minimize the purity and therefore
carry the largest off-diagonal weight allowed.

The remaining question is which off-diagonal terms survive the time average.
In Appendix~\ref{app:jc}, we evaluate this channel, including repeated-gap and
accidental-degeneracy cross terms that can appear in a finite cutoff.  These
extra cross terms do not change the JC value.  The JC dip is therefore not the
spin-chain mechanism of Ref.~\cite{Low:2026ep}, where the dips come from exact
$SU(2)$ multiplet degeneracies at the XXX point or repeated free-fermion gaps
at the XX point.  Instead, Eq.~(\ref{eq:offdiag_weight}) leaves a large
off-diagonal weight, and the $U(1)$ sectors make the nonzero $M_{mn}$ blocks
sparse, with support only between states sharing an oscillator occupation.
Breaking number conservation spreads this weight over many more pairs.
Combining the off-diagonal channel of Eq.~(\ref{eq:jc_off_main}) with the
diagonal channel of Eq.~(\ref{eq:jc_diag_main}) gives
\be
\begin{aligned}
\epb_{\rm JC}(\Ntr)
&=1-\frac{11\Ntr+4}{6\Ntr(\Ntr+1)}\\
&\xrightarrow{\ \Ntr\to\infty\ }\;1
\qquad(\om=\Del).
\end{aligned}
\label{eq:ep_jc}
\ee
For any nonzero $g$, this expression is independent of $g$, because
the eigenvectors in Eq.~(\ref{eq:jc_eigs}) at $\om=\Del$ do not depend on
$g$: the coupling strength changes the level spacings, but does not change the sector structure
or the equal-superposition form of the doublets. At exactly $g=0$ and
$g'=0$, the dynamics is decoupled and the entangling power vanishes, so
Eq.~(\ref{eq:ep_jc}) should be read as the nonzero-coupling JC value.  The
same argument applies to the coherent-state diagnostic, whose value at the
resonant JC point is likewise independent of nonzero $g$.

Figure~\ref{fig:jc} shows the scans across the interpolation.  The finite-Haar
curves meet at the analytic JC value and, for moderate and strong coupling,
the JC point is a local minimum.  At weak coupling, the finite-Haar curve is
nearly flat because the eigenvectors change little over the plotted window.
The coherent diagnostic shows the same common JC value, with a visible dip for
$g/\om=0.5$ and $1.0$ but no resolved weak-coupling dip.

The comparison with the AQRM integer feature is then simple. The AQRM integer point suppresses the diagonal channel and gives a strong peak.
The JC point keeps the diagonal channel near its absolute floor, while the sparse
$U(1)$ sector structure makes the off-diagonal channel large enough to produce
a weak dip.

\begin{figure*}[t!]
\centering
\includegraphics[width=0.92\textwidth]{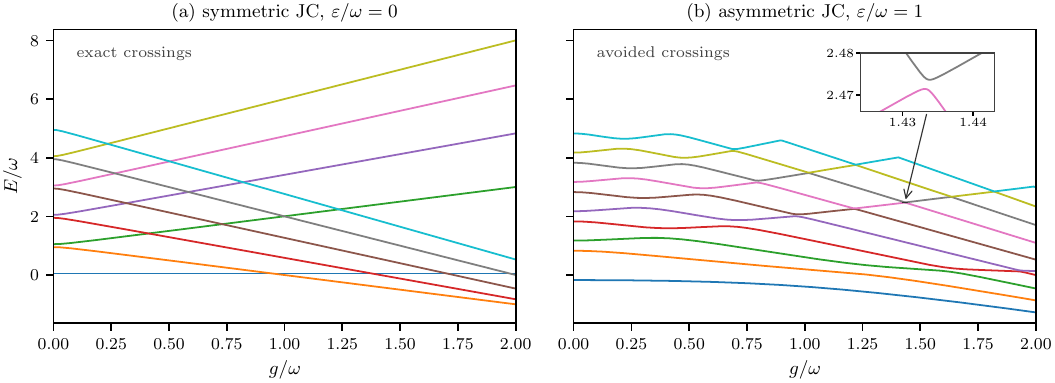}
\caption{Low-lying spectrum of the JC limit of Eq.~(\ref{eq:anisotropic_qrm})
with $g_{\rm cr}=0$ and $\Del/\om=0.9$.  The horizontal axis is $g/\om$, and
in both panels the colors label the ten energy levels that are lowest at
$g=0$. Panel (a) shows the symmetric JC
model, $\ep/\om=0$, where the $U(1)$ symmetry permits
crossings between energy levels.  Panel (b) shows the asymmetric JC model
at the integer bias point $\ep/\om=1$; the same ten energy levels show avoided
crossings rather than exact crossings. The inset in panel (b) magnifies the arrowed
near crossing near $g/\om\simeq1.43$ and shows the finite avoided gap.  We have verified that all the near-crossings in (b) are in fact avoided crossings with small gaps.}
\label{fig:asym_jc_spectrum}
\end{figure*}

It is natural to ask whether the asymmetric JC model, Eq.~(\ref{eq:anisotropic_qrm})
with $g_{\rm cr}=0$, inherits a hidden symmetry at integer values of
$\ep/\om$ similar to the AQRM.  The spectral answer is no.  At $\ep=0$, the JC
Hamiltonian has the manifest excitation-number symmetry discussed above, and
levels belonging to different excitation sectors can cross exactly.  Turning on
the bias term $\ep\sx/2$ breaks this block structure even in the JC limit.  As
shown in Fig.~\ref{fig:asym_jc_spectrum}, the corresponding crossings at
$\ep/\om=1$ are replaced by avoided crossings rather than by a new integer-bias
family of exact degeneracies.  Thus the AQRM hidden symmetry at
$\ep/\om\in{\mathbb Z}$ is not inherited by the asymmetric JC limit.

\section{Conclusion}
\label{sec:discussion}

In this work, we used the time-averaged entangling power as an operator-level
diagnostic of the hidden symmetry in the AQRM, with the manifest $U(1)$
symmetry at the JC point providing a contrast within the same Rabi family.  The
hidden symmetry is not manifest in the Hamiltonian and, at generic fixed
$(g,\Del)$, the integer-bias condition $\ep/\om\in{\mathbb Z}$ does not by
itself produce an exact level crossing.  Thus the diagnostic is not simply
rediscovering a visible conserved charge or a generic spectral degeneracy.
Since the oscillator Hilbert space is infinite-dimensional, we compared two
finite input ensembles: a Haar average after Fock-space truncation and a
coherent-state average at fixed mean occupation $\nbar$.  The two ensembles
weight the oscillator Hilbert space differently, but both exhibit peaks at the
integer-bias values where the hidden symmetry resides.

The absence of a normalizable Haar measure on the full bosonic space makes this
comparison nontrivial.  The finite-Haar diagnostic is therefore defined only
after imposing a Fock-space cutoff, and
that cutoff also fixes the typical input occupation,
$\bar n_{\rm H}=(\Ntr-1)/2$.  Coherent-state sampling at fixed $\nbar$ instead
separates the physical input scale from the numerical cutoff.  The two
diagnostics therefore need not agree pointwise, and indeed their amplitudes
differ.  Their common feature is the location of the peaks at the integer-bias
values, which indicates that the peak locations are controlled by the AQRM
eigenstructure rather than by the detailed input ensemble.

The origin of the peaks is an eigenvector effect.  In the displaced-doublet
analysis of Sec.~\ref{sec:aqrm_peaks}, the pair
$|{\rightarrow},j\rangle_d$ and $|{\leftarrow},j+k\rangle_d$ is described, when
approximately isolated, by a two-level Hamiltonian with diagonal entries
controlled by $\eta_k=\ep-k\om$ and off-diagonal matrix element proportional to
$\Del s_{jk}$.  The resulting mixing angle satisfies
Eq.~(\ref{eq:mixing_angle}) and is maximized when $\eta_k=0$, namely at
$\ep=k\om$, independently of $j$.  Thus finite-$\Del$ mixing reorganizes a
whole family of displaced doublets at the same integer bias, shortens the
reduced-qubit Bloch vectors to the overlap-controlled value in
Eq.~(\ref{eq:overlap_floor}), and suppresses the diagonal contribution to
$\langle\Tr\rho_A^2\rangle_t$.  Since the entangling power is
$2(1-\langle\Tr\rho_A^2\rangle_t)$, this suppression gives a peak.  By
contrast, the JC point shows that maximal mixing by itself does not fix the
sign of the response: there the diagonal channel lies near its absolute floor,
but the $U(1)$ sector structure concentrates the off-diagonal reduced blocks
strongly enough to produce a weak dip.

Several questions remain open.  It would be useful to relate the input scale
needed to resolve the $k$th peak directly to the structure of the degree-$k$
hidden-symmetry operator $J_k$.  It would also be interesting to develop a
closed coherent-ensemble analog of the finite-Haar Bloch-quadrupole diagnostic,
rather than using the finite-Haar diagonal channel as a guide.  Since coherent
states are natural inputs in cavity and circuit QED, the coherent-state
entangling power may provide a useful finite-energy probe of eigenvector
reorganization in light--matter systems.  The present results support the
broader idea that operator-level entanglement diagnostics can respond to
symmetry even when the symmetry is hidden and leaves no generic degeneracy in
the spectrum.

\section*{Acknowledgements}

I.~L. is supported in part
by the U.S. Department of Energy under contracts
No. DE-SC0023522, No. DE-SC0010143, and No.
89243024CSC000002 (QuantISED Program). S.A. is supported by Japan's Ministry of Education, Culture, Sports, Science and Technology (MEXT) Quantum Leap Flagship Program Grant Number JPMXS0120319794. This research was supported in part through the computational
resources and staff contributions provided for the Quest
high performance computing facility at Northwestern
University which is jointly supported by the Office of
the Provost, the Office for Research, and Northwestern
University Information Technology.

We acknowledge the use of Claude  (Anthropic) and GPT/Codex  (OpenAI) for
assistance with computations, writing, and figure generation. The authors take full responsibility
for the correctness and scientific value of the work.

\appendix

\section{Haar Swap Formalism and Frequency Groups}
\label{app:swap_formalism}

This appendix reviews the finite-dimensional Haar formula underlying Eq.~(\ref{eq:ep_haar}) and the frequency-group language used in the main text; a more detailed treatment of this formalism is given in Ref.~\cite{Low:2026ep}.  

Consider two copies of the bipartite Hilbert space, ordered as $A_1B_1A_2B_2$.  We denote by $T_{13}$ the swap between $A_1$ and $A_2$, and by $T_{24}$ the swap between $B_1$ and $B_2$.  The elementary identity is the swap trick,
\be
\Tr \rho_A^2=\Tr\left[(\rho_A\otimes\rho_A)T_{13}\right],
\label{eq:swap_trick}
\ee
where the trace on the right-hand side is over the two $A$ copies.  Applying the unitary before tracing out $B$ gives
\be
\Tr\rho_A^2
=\Tr\left[
U^{\otimes2}\left(|\psi,\phi\>\<\psi,\phi|\right)^{\otimes2}
U^{\dagger\otimes2}T_{13}\right].
\label{eq:rhoA2_twocopy}
\ee
Thus the nonlinear quantity $\Tr\rho_A^2$ becomes a linear trace on a doubled space.

The Haar average can now be evaluated explicitly.  On a $d$-dimensional Hilbert space,
\begin{align}
\int\dd\mu_{\rm H}(\psi)
\left(|\psi\>\<\psi|\right)^{\otimes2}
&=\frac{\Id+S}{d(d+1)}
\equiv C_d(\Id+S),\nonumber\\
C_d&=\frac{1}{d(d+1)} ,
\label{eq:haar_moment_app}
\end{align}
where $S$ swaps the two copies.  For product inputs on $A$ and $B$, Eq.~(\ref{eq:haar_moment_app}) gives
\be
\overline{\left(|\psi,\phi\>\<\psi,\phi|\right)^{\otimes2}}
=C_{d_A}C_{d_B}(\Id+T_{13})(\Id+T_{24}) .
\label{eq:product_haar_moment_app}
\ee
Substituting this expression into Eq.~(\ref{eq:rhoA2_twocopy}) gives the
finite-dimensional Haar average of the reduced purity,
\be
\overline{\Tr\rho_A^2}
=C_{d_A}C_{d_B}\sum_{\alpha=0}^1 I_\alpha(U),
\label{eq:purity_swap_app}
\ee
and therefore, with the normalization of Eq.~(\ref{eq:ep_haar}),
\be
\mathrm{ep}(U)=
\frac{d_A}{d_A-1}
\left[
1-C_{d_A}C_{d_B}\sum_{\alpha=0}^1 I_\alpha(U)
\right].
\label{eq:ep_swap_app}
\ee
where
\be
I_\alpha(U)=
\Tr\left(T_{1+\alpha,3+\alpha}\right)
+\left\<U^{\otimes2},
T_{1+\alpha,3+\alpha}U^{\otimes2}T_{13}
\right\>_{\rm HS}.
\label{eq:Ialpha_app}
\ee
Here $\langle A,B\rangle_{\rm HS}\equiv\Tr(A^\dagger B)$, and $T_{1+\alpha,3+\alpha}$ denotes $T_{13}$ for $\alpha=0$ and $T_{24}$ for $\alpha=1$.  The first term in Eq.~(\ref{eq:Ialpha_app}) is the input-independent swap trace; the second contains the nontrivial dynamics.

We now specialize to $U(t)=\exp(-iHt)$, with $H|E_m\>=E_m|E_m\>$.  Since Eq.~(\ref{eq:Ialpha_app}) is quartic in $U(t)$ and $U^\dagger(t)$, the time dependence appears through phases of the form
\be
\exp\left[-i(E_m-E_n-E_p+E_q)t\right].
\label{eq:quartet_phase_app}
\ee
The infinite-time average therefore keeps only quartets satisfying
\be
E_m-E_n=E_p-E_q .
\label{eq:freq_condition_app}
\ee
It is convenient to group ordered pairs $(m,n)$ by their common energy difference $\Omega=E_m-E_n$.

To write the resulting sums, reshape each eigenvector as a $d_A\times d_B$ coefficient matrix, $|E_m\>=\sum_{a,b}(C_m)_{ab}|a\>_A|b\>_B$, and use the matrices $M_{mn}$ defined in Eq.~(\ref{eq:Mmn_def}).  We also define the $B$-side contraction
\be
\widetilde M_{mn}\equiv C_m^\dagger C_n .
\label{eq:Mtilde_app}
\ee
For the $\alpha=0$ invariant, the dynamical piece is a positive group sum,
\be
I_0^{\rm dyn}
=\sum_\Omega
\sum_{\substack{(m,n)\in g_\Omega\\(p,q)\in g_\Omega}}
\left|\Tr\left(M_{mn}^\dagger M_{pq}\right)\right|^2
=\sum_\Omega \|G_\Omega\|_F^2 ,
\label{eq:I0_group_app}
\ee
where $g_\Omega=\{(m,n):E_m-E_n=\Omega\}$ and $G_\Omega$ is the Gram matrix of the matrices $M_{mn}$ within the group.  The companion invariant has the same frequency grouping,
\be
I_1^{\rm dyn}
=\sum_\Omega
\sum_{\substack{(m,n)\in g_\Omega\\(p,q)\in g_\Omega}}
\Tr\left(M_{mn}^\dagger M_{pq}\right)
\Tr\left(\widetilde M_{mn}^\dagger\widetilde M_{pq}\right),
\label{eq:I1_group_app}
\ee
with the $B$-side contraction carried by $\widetilde M_{mn}$.

Eqs.~(\ref{eq:I0_group_app}) and (\ref{eq:I1_group_app}) make explicit the two
ingredients of the finite-Haar diagnostic.  The spectrum decides which ordered
pairs are grouped together, while the eigenvectors determine the overlaps
inside each group.  Exact degeneracies enlarge the zero-frequency group, and
repeated gaps enlarge nonzero-frequency groups.  In Eq.~(\ref{eq:I0_group_app})
this can add nonnegative cross terms, increasing
$\langle\Tr\rho_A^2\rangle_t$ and pushing the entangling power in
Eq.~(\ref{eq:ep_swap_app}) downward.  The AQRM peaks studied in the main text
are different: at generic fixed $(g,\Del)$ the integer-bias condition does not
enlarge the relevant frequency groups, so the response must come from
eigenvector reorganization.

\section{Fixed-Radius and Gaussian Coherent Sampling}
\label{app:coherent}

\begin{figure}[htbp]
\centering
\includegraphics[width=\columnwidth]{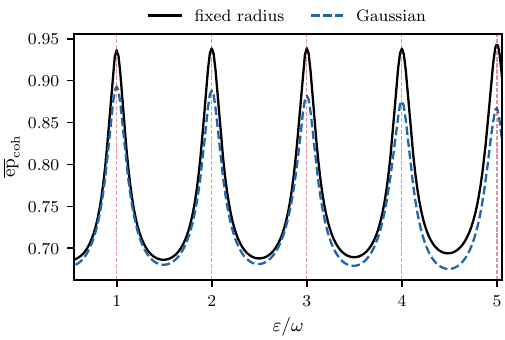}
\caption{Fixed-radius and Gaussian coherent sampling at $\nbar=20$, $g/\om=0.50$, and $\Del/\om=0.5$.  The two ensembles have the same mean occupation.  Both show peaks at the same integer-bias values.}
\label{fig:app_fixed_gaussian}
\end{figure}

The coherent-state diagnostic in the main text uses the fixed-radius measure
in Eq.~(\ref{eq:fixed_radius_mu}), which fixes $|\alpha|^2=\nbar$ and averages
only the phase.  This appendix compares that choice with a second coherent
ensemble at the same mean occupation.  The purpose is to check which features
are tied to the AQRM eigenstructure, rather than to the detailed occupation
weights of the chosen coherent ensemble.

The comparison ensemble is the Gaussian coherent ensemble
\be
\mu_{\nbar}^{\rm G}(\alpha)=\frac{1}{\pi\nbar}\exp\left(-\frac{|\alpha|^2}{\nbar}\right).
\label{eq:gaussian_mu}
\ee
This weights oscillator occupations thermally and is the Glauber--Sudarshan
$P$ representation of a thermal oscillator state of mean occupation
$\nbar$~\cite{Glauber:1963fi,Sudarshan:1963ts}.  The thermal state appears as
the first moment of the ensemble only.  Since the entangling-power average is
quadratic in the output density matrix, the relevant oscillator input object is
the two-copy moment
\be
M_2^{\rm G}(\nbar)=\int\dd^2\alpha\,\mu_{\nbar}^{\rm G}(\alpha)
\left(|\alpha\>\<\alpha|\right)^{\otimes2}.
\label{eq:M2def}
\ee
In the Fock basis, with $\zeta\equiv2+1/\nbar$,
\begin{align}
\<m_1m_2|M_2^{\rm G}(\nbar)|n_1n_2\>
&=
\frac{\delta_{m_1+m_2,n_1+n_2}}{\nbar}
\frac{(m_1+m_2)!}{\zeta^{m_1+m_2+1}}
\nonumber\\
&\quad\times
\frac{1}{\sqrt{m_1!m_2!n_1!n_2!}} .
\label{eq:M2_matrix}
\end{align}
This moment is normalized and block diagonal in total Fock occupation, but it
does not factorize as $\rho_{\rm th}\otimes\rho_{\rm th}$.  The two copies
share the same sampled amplitude $\alpha$, so the moment contains replica
correlations rather than two independent thermal draws.  Its envelope decays
geometrically as $(2/\zeta)^{m_1+m_2}\sim \exp[-(m_1+m_2)/(2\nbar)]$.  Thus, at
fixed $\nbar$, the Fock cutoff is only a convergence parameter once the
coherent tail is contained; this differs from finite-Haar sampling, where
changing $\Ntr$ changes the input ensemble itself.

For the fixed-radius scans, we evaluate the per-input expression in
Eq.~(\ref{eq:Rchi}) and then average over the coherent phase.  For the Gaussian
ensemble, Eq.~(\ref{eq:M2_matrix}) gives the corresponding two-copy oscillator
moment.  We checked the fixed-radius results by increasing $\Ntr$; once the
coherent tail is contained, the curves are unchanged at the plotted precision.

Figure~\ref{fig:app_fixed_gaussian} compares the fixed-radius ensemble with
the Gaussian ensemble in Eq.~(\ref{eq:gaussian_mu}).  The two choices weight
the Fock window differently and therefore need not agree in amplitude, but
both use $\nbar$ rather than $\Ntr$ as the physical input scale.  Their common
feature is the location of the integer-bias peaks.

\section{JC Point: Finite-Haar Evaluation}
\label{app:jc}

This appendix gives the finite-Haar evaluation used in Sec.~\ref{sec:jc}.
We take $d_A=2$ and $d_B=\Ntr$, with
$M_{mn}$ as in Eq.~(\ref{eq:Mmn_def}) and
$\rho_m\equiv\rho_A^{(m)}=M_{mm}$.  Applying the product Haar moment in
Eq.~(\ref{eq:product_haar_moment_app}), the contribution from the diagonal
pairs is
\be
\begin{aligned}
\left\<\Tr\rho_A^2\right\>_t^{\,\Omega=0}
&=\frac{1}{6\Ntr(\Ntr+1)}
\sum_{m,p}\\
&\quad\times\Big[1+\delta_{mp}+\Tr(\rho_m\rho_p)
+\|M_{mp}\|_F^2\Big]\\
&\quad\times\Tr(\rho_m\rho_p) \ ,
\end{aligned}
\label{eq:diag_channel_app}
\ee
For a nonzero-frequency singleton $(m,n)$, paired only with its reverse
$(n,m)$ in the time average, the contribution is
\be
\begin{aligned}
\left\<\Tr\rho_A^2\right\>_t^{\,\Omega\ne0}
&=\frac{1}{6\Ntr(\Ntr+1)}
\sum_{m\ne n}\left\|M_{mn}\right\|_F^2\\
&\quad\times\Big(1+\left\|M_{mn}\right\|_F^2
+\Tr(\rho_m\rho_n)\Big) \ .
\end{aligned}
\label{eq:offdiag_channel}
\ee
At the resonant JC point, $\om=\Del$, $g'=0$, and $g\ne0$, the doublet
eigenstates in Eq.~(\ref{eq:jc_eigs}) have $\rho_m=\tfrac12\Id$.  The two
boundary singlets are pure.  Substituting these reduced states into
Eq.~(\ref{eq:diag_channel_app}) gives
\be
\left\<\Tr\rho_A^2\right\>_t^{\,\Omega=0}
=\frac12+\frac{1}{2\Ntr(\Ntr+1)} \ .
\label{eq:jc_diag_main}
\ee
The off-diagonal sum is fixed by the sparse JC sector structure.  The
nonzero blocks come from pairs within a doublet, pairs between adjacent
doublets sharing one oscillator occupation, and pairs between a boundary
singlet and its adjacent doublet.  Their total contribution to the numerator
of Eq.~(\ref{eq:offdiag_channel}) is
\be
\begin{aligned}
S_{\rm off}
&=2(\Ntr-1)\cdot1
+8(\Ntr-2)\cdot\frac{7}{16}
+8\cdot1\\
&=\frac{11\Ntr-2}{2} \ ,
\end{aligned}
\ee
and therefore
\be
\left\<\Tr\rho_A^2\right\>_t^{\,\Omega\ne0}
=\frac{11\Ntr-2}{12\Ntr(\Ntr+1)} \ .
\label{eq:jc_off_main}
\ee
Adding the two channels gives
\be
\left\<\Tr\rho_A^2\right\>_t=\frac12+\frac{11\Ntr+4}{12\Ntr(\Ntr+1)} \ ,
\ee
and $\epb_{\rm JC}=2(1-\left\<\Tr\rho_A^2\right\>_t)$ reproduces
Eq.~(\ref{eq:ep_jc}).  Repeated or degenerate gap groups that occur in the
finite cutoff do not change this result: the relevant cross terms vanish or
cancel in the grouped evaluation.  We have verified Eq.~(\ref{eq:ep_jc})
against the grouped finite-Haar evaluation for $\Ntr=4,\dots,60$, and checked
at $\Ntr=30$ that the accidental degeneracies at $g=\om$ leave the total
unchanged to machine precision.

\bibliographystyle{apsrev4-2}
\bibliography{refs}

\end{document}